## The *Big Brother Naija* TV Reality Show as Coordinate of Media Functions and Dysfunctions

[1]Bolu John Folayan, PhD, [2]Olubunmi Ajibade, PhD, [3]Olubunmi Dipo-Adedoyin, [4]Toyin Segun Onayinka, PhD, [5]Toluwani Titilola Folayan

[1]Department of Mass Communication, Joseph Ayo Babalola University, Ikeji Arakeji, Osun State, Nigeria

[2]Department of Mass Communication, University of Lagos, Akoka, Yaba, Nigeria

[3]Department of Mass Communication, Lagos State Polytechnic,, Ikorodu, Nigeria

[4]Department of Mass Communicaiton, Federal University, Oye-Ekiti, Nigeria

[5]Department of Mass Communication, Redeemer's University, Ede, Nigeria

[1]bolujohnfolayan@gmail.com, [2]bunmiajii@yahoo.ca,, [3]bunmiddoyin21471@gmail.com

[4]toyin.onayinka@fuoye.edu.ng , [5]toluwanifolayan@gmail.com

**ABSTRACT**

The mass media play at least five basic functions which include news dissemination, surveillance of the environment, correlation of the components of the society, entertainment and transmission of social heritage. Sometimes, disruptions and impairments do occur in the performance of these roles and some of these basic functions become dysfunctions, which turn the media into purveyor of negative values. The present study investigates how popular the Nigerian TV reality show, Big Brother Naija (BBN), is perceived by its viewers. Three hundred heavy viewers of the programme were surveyed from Lagos and Ede, South-West Nigeria, and their opinions and attitudes were sought regarding; why they like or dislike the programme; the gratifications that those who like the programme derive and whether the BBN, as media content, is generally functional or dysfunctional to the society. Sixty-six per cent 66 (33.7%) of respondents like the programme because it entertains. Half of the respondents, 99(50.5%) dislike 'immoral aspects' of the programme. The viewers affirm that the eviction part of the programme was their highest form of gratification. Most respondents, despite public outcry against the programme, consider the programme to be "functional". Findings reinforce the postulation that TV viewers are not passive consumers of media contents.

## INTRODUCTION

The mass media (newspapers, magazines, books, periodicals, radio, television, films and the social media) perform key roles in the development of a society such as serving as forum for debate, political participation, re-orientation, socialization, mobilization, education and entertainment. (Lasswell, 2013). Of course, the media also serve as platform that connect advertisers with consumers and potential consumers (advertising function). Each mass media type varies in the effectiveness with which it performs the foregoing functions. According to Moemeka, a single communication message is capable of performing different functions (eliciting different consequences) in society- manifest and latent functions and dysfunctions. The manifest function connotes the media achieving its intended purpose; the latent function refers to the unintended positive outcome the message achieves while dysfunction refers to the inability of a message to achieve its intended purpose and doing something directly against or opposite the intention of the message (Moemeka, 2016).

The functions of the mass media inter-relate with the consequence or impact of the message, which is mainly determinant by the predisposition of the audience and contexts in which they perform their duties. Each function has an element in another function or are a result of an attempt to reduce the dysfunction of the media (Moemeka 2016). In this sense, dysfunctions allude unproductive role the media serve in society or its ineffectiveness in performing their roles. For example, experts have suggested that television tends to corrupt its viewers, especially if what is shown is negative and that it wastes productive time and energy (Gerber Gross,

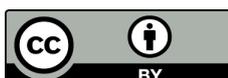

1976). Media effects are the intentional or unintentional reverberations or outcome of what the media do (McQuail, 2010).

This study is an investigation into the functional or dysfunctional role of the mass media – using broadcasting (television network) and a popular Nigerian reality TV show, *Big Brother Naija* as a case study.

### *History of Reality Shows in Nigeria*

Reality shows gained popularity in Nigeria in 2003, when Bayo Okoh, a Nigerian, featured in the first season of *Big Brother Africa* (BBA) which is an adaptation of the international Big Brother series. BBA is a reality show which showcases 12 housemates from 12 different African countries living in an enclosed house for a stipulated period. The winner usually ends up with a cash prize coupled with fame. It, however, stopped transmission after nine seasons (Leviton, 2018).

It was widely speculated that the reason for BBA's popularity in Nigeria was the fact that Nigerians were eager and excited to be represented in an international/African programme as it was the first time that the reality show would be created using participants of different personalities from a continent. The relative success of *Big Brother Africa* with Nigerians resulted in another local reality show called *Gulder Ultimate Search* (GUS). The success of GUS also paved the way for more reality programmes such as *MTN Project Fame West Africa*, *Glo's Naija Sings*, *Maltina Dance Hall*, *Star Quest* produced by Nigerian Breweries and *Big Brother Naija*. (Gonnym, 2018)

The first season of BBN was launched on March 5, 2016 and lasted for 91 days. Subsequent editions were held in 2017, 2018, 2019 and 2020. The fourth season, tagged 'Double Wahala' recorded unprecedented viewership and ratings as over 170 million votes were recorded from viewers (Izuzu, 2018).

Despite its soaring popularity, a large number of people have criticized the *Big Brother Naija* programme, arguing that it depicts indecent acts such as nudity, vulgarism, alcoholism, sex, among others (Anazia, 2018).

### *Statement of the Problem*

Scholars hold different perspectives on the utilitarian value of media contents. On one side of the divide are development communication practitioners who posit that the media ought to publish only contents that could facilitate the development of the society (for example, Nora C. Quebral, Denis McQuail, and Frank Ugboajah). On the other side are those who promote the reflective theory – the media can only serve as mirrors of what happens in the society as both so-called negative and positive contents play important roles in shaping the society; (for example, Michael Gurevitch and Lee Loevinger).

The foregoing perspectives technically suggest that media contents and media audience exert influences. The extent to which media-audience interactivity leads to specific outcomes has thus generated tremendous research among scholars. There remains a gap in such studies in Nigeria. For example, why would a highly-criticized programme such as BBN continue to generate heavy viewership, considering that these criticisms also emanate from the audience of television? The excerpt below is typical of series of criticisms against BBN:

The on-going *(sic)* Big Brother Nigeria (*BBNaija*) has stolen the hearts of most Nigerians, especially the youth. It has created and still creates a buzz, particularly in the social media space, with *(sic)* high level of engagement in voting and followership among the Nigerian youth in particular, who devote their time and resources to follow the show religiously throughout the 85 days it features on MultiChoice's *DStv* and *GOtv* platforms.

Besides its entertainment value, the show has raised several cogent debates about morality and the essence of hard work, as the winner walks away with ₦25 million prize [more than $55,000] and ₦20 million [more than $44,000] worth of SUV, amounting to a total package of ₦45 million. Investigations revealed that during the (Season 2, See Gobe), 11 million votes were recorded at the last eviction before the grand finale. This increased as the frenetic campaigns on social media got heightened, putting the final vote tally at over 26 million, which was almost the 28.5 million votes recorded in the 2015 general election (Guardian.ng, 2018).

From the foregoing, it is expedient to offer scholarly insight into the debate on the appropriateness of the programme.

*Research Questions*

1.	Why do viewers like or dislike the BBN programme?

2.	What gratifications do viewers who like the programme derive from the programme?

3.	To what extent is the BBN, functional or dysfunctional to the society?

**LITERATURE REVIEW**

*Conceptual Framework*

*Meaning and impact of reality TV programmes*

Kilborn (1994) states that three criteria unite to comprehensively define what entails reality programming. These are:

- Recording 'on the wing' and frequent events in the lives of individuals and groups through the aid of lightweight video equipment.
- Attempt to simulate real-life events through various forms of dramatized reconstruction;
- The incorporation of the recording in suitably edited form into an attractively packaged television programme which can be sponsored on the strength of its reality credentials.

Wei and Tootle (2002:6) define reality television as

TV shows that simulate real-world, real-life psychologically, mentally or emotionally challenging situations, involving reward-motivated, self-selected contestants from the audience. The contestants act spontaneously, improvise, and showcase their real emotions in meeting the challenges they encounter in real settings.

In reality programming, non-actors in unscripted situations act as contestants and behave spontaneously with some level of producer-involvement. Experiences on the programme are comprehensively captured with little or no restrictions to particular moments (Gardyn, 2001).

Many studies suggest that reality television shows make personal thoughts, behaviour and interactions of their characters the main hub of the audience's attention. In many countries, reality programmes have been criticized for not being an accurate account of reality. For instance, in 2012, Mike Fleiss, the creator and executive producer of "*The Bachelor*"; a reality TV show, told the "*Today Show*" that 70 to 80 per cent of what people see on reality television is fake.

They're loosely scripted. Things are planted. Things are salted into the environment so things seem more shocking. What we're seeing isn't actually real. It's dramatized reality where contestants are goaded into the most dramatic reactions, and story lines are set up well in advance (Hines, 2012).

Philip Ross of *International Science Times* posits that reality television has a negative influence on viewers' world, basing his opinion on a study by the University of Wisconsin. In the study, 145 students from the university were surveyed based on reality television consumption. The study concluded that reality television viewers have the opinion that the argumentative and conniving behaviour portrayed on television shows is considered normal in today's society (Isciencetime.com, 2018).

Based on another study led by Lisa K. Lundy in 2008, researchers garnered 34 participants and categorized them into four groups to canvass the social effects of reality television. Most of the participants viewed reality television as an escape from reality and an immoral or irresistible phenomenon. In the study, Ross, Lundy & Riccio concluded that reality TV induces problematic behaviours in people, especially towards the younger generation (Isciencetime.com, 2018).

More recent research further shows that reality television is an addicting phenomenon, and analysts have claimed that society is so addicted to shows, such as *American Idol, Keeping Up With the Kardashians* and *The Real World*, because it is interactive, entertaining and relatable with its audiences. Ogunade (2018) in her study on "*Big Brother Naija and Brand Positioning among OAU students*" also examined *Big Brother Naija* and the brand behind its sponsorship. The findings of the research enlightened the positioning of the brand and its importance as it showed that programme had effectively positioned the brand behind its sponsorship. The study also revealed that a high percentage of the respondents (90.83%) watch the *Big Brother Naija* reality TV show while 9.17% do not watch the reality TV show.

In her earlier study of *Big Brother Naija)*, Lwahas found that reality television has made a grand entrance into the landscape of programming in the twenty first century. This, according to her, is evidenced by the fact that reality television phenomenon has become the mainstream of television programming, providing relatively cheap entertainment (using ordinary people, no scripts and replicated format).

Based on the foregoing studies, observations, and views from researchers, the connection between reality television and societal behaviour is vague. But the growing popularity of Reality TV shows suggests that its perceived dysfunctions can be re-directed or adapted to positive ends (Murray, 2010).

***Theoretical framework***

The theoretical frame for this study rests on a tripod that describes the historical phases of media uses and impact: The Bullet/Hypodermic Paradigm; the Uses and Gratification Paradigm and the Reflective-Projective Paradigm.

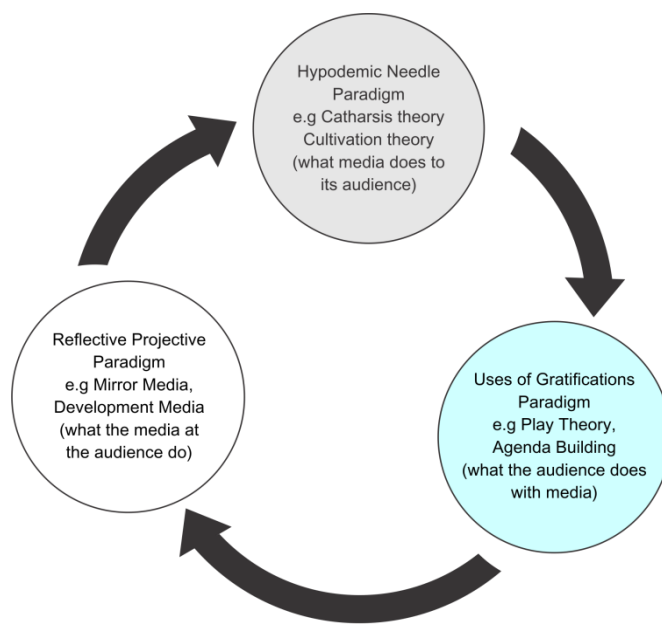

*Fig 1. A theoretical frame of how the media and its audience shape reality*

I*The Bullet or Hypodermic Paradigm*: Theory and research under this paradigm affirmed that television effect directly impacts the audience. The submissions of Gerbner and Gross (1976), succinctly represent this perspective. According to the paradigm, television is the prevailing force in impacting the society and the messages conveyed through television reach large, heterogeneous audiences that traditional media were unable to. Audience consumption of television content is ritualistic, and not based on viewers watching some particular programmes; that is, viewers watch more by the clock than by the programme (Gerbner, *et al.*, 1986). Individual tastes and p**references** in television content are not an important factor in cultivation research, or at least not as imperative a factor as viewing patterns associated with time. Gerbner argues that people who are regularly exposed to media for long periods are more likely to perceive the world's social realities as they are presented by the media they consume which in turn affects their attitudes and behaviours. This cognitive bias (known as the *mean world syndrome)* suggests that people may perceive the world to be more dangerous than it is (Bandura, 2009). In short, the theory assumes that mass media messages have immediate, uniform and direct effect on a passive audience.

Many other theories share the foregoing perspective of 'direct impact'. For instance, the Catharsis Theory posits that watching violent media reduces aggression, pity, fear and anger, as viewers can through media exposure purge such feelings. (Florea, 2013).

II. *Uses and Gratifications Paradigm:* The uses and gratifications (U&G) approach has a long-standing history in communication research and mass communication. Fundamental to the theory is the aim to understand how, why, and with what purpose people use the media in their everyday lives (Liu, 2015).

Uses and gratification theory explains why people seek out the media (content) and what they use it for. UGT assumes that media consumers have control over their media usage. They are not passive consumers of media products. Rather, they intentionally seek out the media to fulfil their individual needs such as socializing, relaxation and entertainment. Thus, through UGT, communication researchers have been able to understand how media audience connects with media content and the web of mediating factors. (Larose & Easton, 2010).

William Stephenson's postulation in 1967 fundamentally provided theoretical and heuristic grounding for the avalanche of UGT studies in the 1970s. Stephenson argues that people readily delineate imagery from real events. He divided human activity into work and play. "Work" refers to 'reality' or 'production' while "play" has to do with relaxation, self-satisfaction and entertainment. Stephenson posited that people use communication as play, much more than as work – more for pleasure and entertainment (Stephenson, 1967).

Numerous scholars have provided further insight on UGT (for example, West and Turner, 2004; Blumler, 1979; Katz, Gurevitch, and Haas, 1973; Rubin, 1983 and Ruggiero, 2009).

West and Turner provided a simple but all-inclusive typology of the postulations of the Uses and Gratifications Theory.

- The audience is active and its media use is goal oriented.
- People have variegated uses (needs) they seek to attain through media.
- Audience members take initiative to connect need gratification to a specific media.
- The media compete with other sources for need satisfaction.
- People have enough self-awareness of their own media use, interests, and motives to be able to provide researchers with a precise picture of that use.
- Value judgments of media content can only be evaluated by the audience.
- (West and Turner, 2004)
- The Audience is said to be active when it has the following under-listed meanings associated with it**:**
- Utility: media have uses for people and people can put media to those uses
- Intentionality: the initial motivations people have determine their consumption of media content.
- Selectivity: individual audience members' use of media may reflect their existing interests and preference.
- Imperviousness to Influence: media audience construct their own interpretation from content that then affects what they think and do. They can avoid certain types of media influence (Blumler, 1979).
- Katz, Gurevitch, and Haas (1973) propose the following as Need Categories:
- Cognitive (acquiring information, knowledge)
- Affective (emotional, pleasurable feelings)
- Personal Integrative (enhaning credibility, status)
- Social Integrative (interacting with family and friends)
- Tension Release (escape and diversion)

Rubin (1983) makes secondary analysis of television viewing motivations and patterns and located two television user types: (1) users of the television medium for time consumption and entertainment; and (2) users of television content for non-escapist, information. The motivations for viewing television comprise of the following: companionship; to pass time; for excitement; for enjoyment; to escape; for relaxation for information; to learn about specific content.

Ruggiero argues that computer-mediated communication has further expanded perspectives of behavioural modes under UGT (Ruggiero, 2009).

III. *The Reflective-Projective Paradigm:* Criticisms of early U&G research centre on the fact that it: (a) relied heavily on self-reports, (b) was unsophisticated about the social origin of the needs that audiences bring to the media, (c) was too uncritical of the possible dysfunction both for self and society of certain kinds of audience satisfaction, and (d) was too captivated by the inventive diversity of audiences used to pay attention to the constraints (Ruggiero, 2000). Many researchers, therefore revisited the Reflective-Projective Paradigm with theoretical underpinnings from Lee Loevinger's reflective theory (also known as the 'linkage theory')

Loevinger submits that the media essentially reflect the inherent social structures of the society. The films, newspapers, books and magazines that people are exposed to reflect the values, norms and beliefs of the people. In other words, the media merely mirrors the society but in this case, the mirror they represent is an ambiguous one. While the media mirrors the society in an organized form, individual audience members project their own individual reflections into the images presented.

But how do the media "merely" serve as reflectors of what happens in the society? Are mirrors extensions of nature or do they hold their symbolic form? If the audience brings individual experiences to bear in interpreting what is reflected in the mirror, then it can be asserted that the rose that is reflected in the mirror is not an *actual* rose itself but a reflection of it, and what each person sees in the mirror is, in reality, not the same (Fishman, 2014).

The theoretical underpinning from the above tripod is that both the media and the audience jointly define the impact or outcome of a communication experience. Each leg of the tripod cannot be ignored depending on context, content, media type and several other factors. For example, the 'hypodermic effect' still holds in 'some kinds of situations' such as internet-mediated communication, just as the reflective paradigm does in other situations. Berger (1998) has provided a classic illustration of the foregoing in explaining how both the media and audience both determine the outcome of communication in cinematography as follows:

**Table 1: Berger's signifier and signified typology in cinematography impact**

| Signifier (shot) | Definition | Signified (interpretation) |
|---|---|---|
| Close-up | Face only | Intimacy |
| Medium shot | Most of the body | Personal relationship |
| Long shot | Setting and characters | Context, scope, public distance |
| Full shot | Full body of person | Social relationship |

**METHODOLOGY**

The survey method was adopted for this study to effectively discover the attitudes, perceptions, thoughts and dispositions of respondents on the subject of study (Wimmer & Dominick, 2011). The research instruments deployed were questionnaire and interview schedule.

The purposive sampling technique was used to select 300 heavy viewers of the BBN in Lagos (the entertainment hub of the country) and Ede (a sub-urban city in Osun State). Lagos was the colonial and political capital of Nigeria and is currently her economic capital, hosting the largest assemblage of Nigeria's over 350 ethnic groups. Ede is the fourth largest town in the State of Osun. It has three tertiary institutions and is about ten kilometres to Osogbo, the capital of the State of Osun. To re-qualify for sampling, prospective respondents must have watched regularly at least the last two of the most recent three editions of the programme. The researchers also ensured that the sampling frame was gender-sensitive. To give perspectives to the responses to the questionnaire, interviews were conducted with two mass communication experts on the objectives of the study. The interviewees were professors of mass communication with Redeemer's University, Ede, and the Lagos State University, Ojoo, Lagos

**RESULTS/FINDINGS**

**I. Why viewers like or dislike BBN**

***Bio-data of respondents***

**Table 2: Age of respondents**

| Age (Years) | Frequency (Percentage) |
|---|---|
| Less than 20 | 87 (29%) |
| 21-26 | 145 (48.3%) |
| 27-32 | 27 (9%) |
| Above 32 | 41 (13.7%) |
| Total | 300 (100%) |

Chi-square= 220.014.

Table 2 shows that the chi-square value of 220.014 with a value <0.05 significant level. This test is significant, as it implies that more people less than 27 years have more interest in the programme.

**Table 3: Educational qualification of respondents**

| Highest educational certificate | Frequency(Percentage) |
|---|---|
| Primary school certificate | 31 (10.5%) |
| Secondary school certificate | 164 (54.7%) |
| Tertiary education | 75( 25%) |
| No formal education | 30(10%) |
| Total | 300 (100%) |

The chi-square value = 158.1147

*Note: There were 52 undergraduate respondents among the respondents classified under "school certificate holders" because that was their highest educational qualification at the time of the study.*

The chi-square value with 3-degree of freedom = 158.1147, p-value 0.000. Table 2, therefore, suggests that the most viewers of BBN (significance value of <0.05) are secondary school certificate holders or were undergraduates.

**Table 4: Gender composition of respondents**

| Sex | Frequency (Percentage |
|---|---|
| Male | 136 (45.3%) |
| Female | 164 (54.7%) |
| Total | 300 (100%) |

As presented in Table 4, female respondents were slightly more than male respondents. This does not necessarily suggest that the female gender watched the programme more as it might have resulted from sampling bias.

*Likeness for BBN*

**Table 5: Like versus dislike of the BBN programme (generally)**

| Viewers' disposition | Frequency (Percentage) |
|---|---|
| Like BBN | 210 (70.1%) |
| Dislike BBN | 65 (21.6%) |

| Can't say categorically | 25 (8.3%) |
|---|---|
| **Total** | **300 (100%)** |

One of the three objectives of this investigation was to determine the extent of likeness and dislike by viewers for the BBN programme. An overwhelming percentage of viewers (70.1 per cent) expressed likeness for the programme generally, even though they have a few things that they would prefer not to be in the programme. About one-tenth of the respondents could not say categorically if they liked or disliked the programme. (See Table 5). Table 6 presents details of the most exciting aspects of the programme to viewers, showing that the 'eviction' session is the most exciting to the viewers.

**Table 6: Aspect of BBN Programme most exciting**

| **Most exciting aspect** | **Frequency (Percentage)** |
|---|---|
| Eviction session | 89 (29.6%) |
| Diary session | 68 (22.6%) |
| Games session | 62 (20.6%) |
| Other | 81 (27.2%) |
| **Total** | **300 (100%)** |

Against the backdrop of heavy criticisms against the BBN in the media, the researchers sought to know why viewers liked the programme. From Tables 7 and 8, it is clear that the viewers loved the programme because it is "entertaining" (40.9 per cent); "educative" (28.5 per cent); creative (15.7 per cent). "Entertaining" in this sense means the programme is amusing. One-tenth of respondents attributed their likeness for the programme to contents which promote Nigerian culture. On the other hand, more than half of the respondents disliked the programme due to its "immoral" contents and because the programme is time-wasting.

**Table 7: Reasons BBN is liked by viewers**

| **Reasons for likeness** | **Frequency (Percentage** |
|---|---|
| It is entertaining | 86 (40.9%) |
| It is original/creative | 33 (15.7%) |
| It is educative | 60 (28.5%) |
| It promotes Nigerian culture | 23 (10.9%) |
| Can't say | 8 (3.8%) |
| **Total** | **210 (100%)** |

**Table 8: Reasons BBN is disliked by viewers**

| **Reasons for dislike** | **Frequency (Percentage** |
|---|---|
| It condones immorality | 34 (37.7%) |
| It is time-wasting | 20 (22.2%) |
| It is not real | 13 (14.4%) |
| It does not add value to viewer | 13 (15.5%) |
| Can't say | 9 (10.2%) |
| **Total** | **90 (100%)** |

**II. Gratifications derived from BBN**

Four out even ten respondents suggested "leisure/relaxation" as their major gratification from watching BBN. (Table 9). The table further shows that "knowledge/Learning" came next to leisure/relaxation in order of gratification-seeking by viewers. Less than ten per cent of respondents watched the programme just to conform to what others were doing while 7.7 per cent used the programme to connect their screen idols.

**Table 9: Perceived benefits (gratifications) of BBN viewers**

| Gratifications | Frequency (Percentage) |
|---|---|
| Leisure/Relaxation | 121 (40.3%) |
| Identification/"Everyone does" | 25 (8.3%) |
| Knowledge/Learning | 28 (9.3%) |
| Entertainment/Fun | 50 (16.7%) |
| Prestige | 25 (8.3%) |
| Connect with my role model | 23 (7.7%) |
| Escape | 18 (6%) |
| Other/Don't know | 10 (3.3%) |
| **Total** | **300 (100%)** |

**III: BBN benefits to viewers and society.**

The present study attempted to determine if the BBN was functional or dysfunctional to the society. Therefore, the researchers explored viewer-perceptions of the programme, the likely reasons for their perception and their opinions regarding commendations and criticism of the programme in the media. Responses are summarized in Table 10 reflecting that two-third of viewers see the programme as functional rather than being dysfunctional. Majority of respondents opposed suggestions that the programme be banned. (Table 11).

**Table 10: Viewers' perception of BBN as functional or dysfunctional for society**

| Perception of viewers | Frequency |
|---|---|
| BBN is functional | 189 (63%) |
| BBN is dysfunctional | 111 (37%) |
| **Total** | **300 (100%)** |

**Table 11: Respondents' views on whether BBN should be banned**

| Perception of viewers | Frequency (Percentage) |
|---|---|
| Ban BBN | 73 (24.3%) |
| Retain BBN | 121 (40.3%) |
| Retain BBN with improvements | 103 (34.3%) |
| Missing item | 3 (1.1%) |
| **Total** | **300 (100%)** |

**The outcome of in-depth interviews with media experts.**

The interviews with two TV broadcasting experts corroborated the foregoing findings. . The experts asserted that the BBN is not dysfunctional to the society. One of them put it succinctly:

From the professional viewpoint, the BBN is not dysfunctional. There are various levels of control that have been undertaken by the regulators and the producers to ensure this is so. For example, the programme is not free-to-air. The viewer subscribes voluntarily, and they are adults. Against sex-related or romantic scenes are edited online or viewers do not have access to them. More educational contents have been infused into the programme as well.

**DISCUSSION/CONCLUSION**

This investigation adds evidence to research literature that suggests that the audience of mass communication impacts communication content just as communication content impacts them. Television viewers are not dormant or passive viewers. Through heavy viewing, participation in voting and chats on various social media, BBN viewers practically impact and define the success of the programme. This finding further reveals the huge potential of the edutainment model for development and marketing communications, especially when media content is made participatory.

Conclusively, data generated from this study show that BBN is more of a functional than a dysfunctional programme. The programme is overwhelmingly perceived by its viewers as beneficial to them and society. In terms of likes and dislikes, the female gender is more than that of the male, with most viewers falling between the ages of 16 and 27 years. Also, most viewers like *Big Brother Naija* because it entertains and helps them to escape boredom. Most of those who dislike the claimed it condones immorality.

Organizers of the programme need to reduce immorality scenes in the programme while infusing more of constructive, entertaining, educative and developmental themes and scenes. Most viewers did not indicate the romantic scenes as attractive to them; hence "offensive sex-related scenes" could be removed to make the programme more popular. As Folayan, Babalola & Abati (2020) have noted, a popular programme perceived as negative has good potential for making positive impact; it only requires strategic reconstruction of its contents. The huge and loyal audience of BBN is a potential asset for companies and marketing communications firms to take advantage of for effective reach and impact.

Lwahas has also reinforced the foregoing by observing thus:

Locally, the adaptation or cloning of television programmes promotes small and medium business development. There will be recognition of innovation, creativity and excellence by producers who want to provide local content to available markets both at international, regional and sub-regional levels. What poses a threat to this attempt in Africa are issues of poor infrastructure, limited capital and restrictive broadcast laws which do not advocate for competition in the international market (Lwahas, 2017).

**REFERENCES**

1. Anazia, D. (2018, March 24). The moral, economic values of reality TV shows. Retrieved January 20, 2019, from Guardian Saturday Magazine: https://guardian.ng/Saturday- magazine/the moral-economic-values of reality-television shows

2. Arjayyay. (2018, November 12). *Nineteen Eighty-Four- Wikipedia*. Retrieved November 15, 2018, from Wikipedia: https://en.m.wikipedia.org›wiki›Ninet...

3. Bandura, A. (2009). Social cognitive theory of mass communication. *Media Psychology.* November.

4. Berger, A. (1998). Media analysis techniques. (2nd edition). California: Sage Publications.

5. Blumler, J.G. (1979). The role of theory in uses and gratifications studies. *Communication Research,* 6. 9-36

6. Fishman, N. (2014). "Through, at, into the looking glass: An argument for mirrors as media." Seminar paper, Princeton University, NJ: USA.

7. Florea, M. (2013). Media violence and the cathartic effect. *Procedia – Social & Behavioural Sciences*. 92. 349-353; *www.sciencedirect.com*

8. Folayan, B.J., Babalola, A.O., & Abati, M.O., (2020). Gratifications, national identity as strategic marketing variables and other promotional strategies in Big Brother Naija TV Programme. *Babcock Journal of Mass Communication,* BJMC). VOL. 5(2), April.

9. Gardyn, G. (2001, September 5). The tribe has spoken: Reality TV is here to stay. *American Demographics*. Retrieved on June 15, 2004 from http://www2.realitytvfans.com/newspub/story.cfm?id=3335.

10. Gerbner, G. & Gross, L. (1976).Living with television: The violence Profile. *Journal of Communication, 26*(2), 173-199.

11. Gerbner, G., Gross, L., Morgan, M., & Signorielli, N. (1986). Living with television: The dynamics of the cultivation process. In J. Bryant & D. Zillmann (Eds.), *Perspectives on Media Effects*. (pp. 17-40). Hillsdale, NJ: Lawrence Erlbaum Associates.

12. Gonnym. (2018, August). *Big Brother Naija- Wikipedia*. Retrieved October 19, 2018, from Wikipedia: https://en.m.wikipedia.org> wiki >Big_

13. Guardian.ng, (2018). https://guardian.ng/saturday-magazine/the-moral-economic-values-of-reality-television-shows)

14. Hines, R. (2012, June 15). Bachelor creator claims 70 to 80 percent reality TV is fake. Retrieved January 24, 2019 from www.today.com : https://www.today.com>popculture>…

15. Isciencetime.com (2018).<http://www.isciencetimes.com/articles/6069/20130916/reality-tv-s-impact-viewers-shows-real.htm, retrieved, October.

16. Izuzu, C. (2018, 3 12). *Head of house, immunity, Big Brother Naija- movies- pulse.ng*. Retrieved 11 6, 2018, from Pulse.ng: https://www.pulse.ng

17. Katz, E., Gurevitch, M., & Haas, H. (1973). On the use of the mass media for important things. *American Sociological Review*, 38, 164-181

18. Kilborn, R. (1994). How real can you get: Recent developments in reality television. *European Journal of Communication, 9*(4), 421-439.

19. Larose, R. & Easton, M.S. (2010). A social cogntive theory of internet uses and gratifications: Towards a new model of media attendane. *Journal of Broadcasting & Electronic Media.* June.

20. Lasswell, H. (2013, January 17). *Functional Theory- Slideshare*. Retrieved September 3, 2018, from LinkedIn Corporation : https://www.slideshare.net>jpbookworld

21. Leviton, I. (2018, September 1). *Big Brother Africa- Wikipedia*. Retrieved 11 6, 2018, from Wikipedia: https://en.m.wikipedia.org>wiki>Big

22. Liu. (2015). A historical overview of Uses and Gratifications Theory. *Cross-cultural Communication*. Vol. 11, No. 9

23. Loevinger, L. (1968). The ambiguous mirror: The reflective-projective theory of broadcasting and mass communications. *Journal of Broadcasting, Vol 12(2).* Published online May 18, 2009.

24. Lwahas, Sarah (2017). Adaptations of reality television programmes: "The Big Brother AfricaReality Show". *Researchjournali's Journal of Media Studies.*Vol. 3 | No. 4. July

25. McQuail, D. (2010). *McQuail's mass communication theory*. London: Sage Publications.

26. Moeemeka, A. A. (2016). *Reporter's Handbook An Introduction to Mass Communication and Effective Journalism*. Abraka Delta: Journal of Communication and Media Research.

27. Murray, S. (2010). "The politics of reality TV: An overview of recent research." In *Media & Society*. 5th ed. Edited by James Curran and Michael Gurevitch, 321–335. London: Bloomsbury Academic, 2010.

28. Ogunade ,T. (2018). "Big Brother Naija and brand positioning among OAU students". A Research Project submitted to the department of Mass Communication. Redeemer's University, Ede Osun State in partial fulfilment for the award for M.Sc in Mass Communication.

29. Rubin, A. M. (1983). An examination of television viewing motivations. *Communication Research*, 8, 141-165.

30. Rubin, A.M. (2009). Television uses and gratifications theory in the 21$^{st}$ century. *Mass Communication and Society.* June

31. Ruggiero, T. E. (2009). Uses and gratifications theory in the 21$^{st}$ century. *Mass communication and Society.* 3(1), 7-32.

*32.* Rutenberg, J. (2001, February 12). Reality shows set off fight over awards. *The New Yorkm Times*, p. C1.

33. Stephenson, W. (1967). The play theory of mass communication. Chicago: University of Chicago Press. Published on line in 1988 by New Brunswick, NJ: Transaction Books.

34. Wei, R. & Tootle, C. (2002). "Gratifications of reality viewing: Antecedents and consequences". Paper presented at the annual convention of the Association for Education in Journalism and Mass Communication, Miami Beach, FL, August 2002.

35. West, R. and Turner, L.H. (2004). Introducing communication theory: Analysis and application, 2$^{nd}$ed. New York: McGraw-Hill.

36. Wimmer, R.D and Dominick, Jr. (2011).Mass media research: An introduction. California: Wadsworth Publishing Company.